\documentclass[a4paper,sort&compress,oneside,preprint,12pt,number]{elsarticle}

\usepackage{amssymb,amsmath,amsthm,mathtools}
\usepackage{subfigure,overpic}
\usepackage{hyperref}
\usepackage{verbatim}
\usepackage{xcolor}

\usepackage[left=2cm,right=2cm,top=2cm,bottom=2cm]{geometry}
\usepackage{placeins}

\DeclareMathOperator{\sech}{sech}

\journal{?}

\begin{document}

\begin{frontmatter}

\title{On the Fractional Dynamics of Kinks in sine-Gordon Models}

  \author[UP]{Tassos Bountis}
      \ead{tassosbountis@gmail.com}

    \author[UHU]{Julia Cantis\'an}
       \ead{julia.cantisan@dci.uhu.es }

    \author[US,IMUS]{Jes\'us Cuevas--Maraver}
    \ead{jcuevas@us.es}

    \author[TU,UAA]{J. E. Mac\'{\i}as-D\'{\i}az}
      \ead{jorgmd@tlu.ee, jemacias@correo.uaa.mx}

    \author[UMASS]{Panayotis G. Kevrekidis}
    \ead{kevrekid@umass.edu}

    \address[UP]{Department of Mathematics, University of Patras, 26500 Patras, Greece}

    \address[UHU]{Grupo de F\'{i}sica No Lineal (FQM-280), Departamento de Ciencias Integradas y Centro de Estudios Avanzados en Física, Matemáticas y Computación, Universidad de Huelva, 21071-Huelva, Spain}
    \address[US]{Grupo de F\'{i}sica No Lineal (FQM-280), Departamento de F\'{i}sica Aplicada I, Universidad de Sevilla. Escuela Polit\'{e}cnica Superior, C/ Virgen de \'{A}frica, 7, 41011-Sevilla, Spain}
    \address[IMUS]{Instituto de Matem\'{a}ticas de la Universidad de Sevilla (IMUS). Edificio Celestino Mutis. Avda. Reina Mercedes s/n, 41012-Sevilla, Spain}

    \address[TU]{Department of Mathematics and Didactics of Mathematics, School of Digital Technologies, Tallinn University, \\ Narva Rd. 25, 10120 Tallinn, Estonia}
    \address[UAA]{Departamento de Matem\'{a}ticas y F\'{\i}sica, Universidad Aut\'{o}noma de Aguascalientes, \\ Avenida Universidad 940, Ciudad Universitaria, Aguascalientes, Ags. 20131, Mexico}

    \address[UMASS]{Department of Mathematics and Statistics, University of Massachusetts Amherst, Amherst, MA 01003-4515, USA}

\begin{abstract}
In the present work we explore the dynamics of single kinks, kink-anti-kink pairs and bound states in the prototypical fractional Klein-Gordon example of the sine-Gordon equation. In particular, we modify the order $\beta$ of the temporal derivative to that of a Caputo fractional type and find that, for $1<\beta<2$, this imposes a dissipative dynamical behavior on the coherent structures.
We also examine the variation of a fractional Riesz order $\alpha$ on the spatial derivative. Here, depending on whether this order is below or above the harmonic value $\alpha=2$, we find, respectively, monotonically attracting kinks, or non-monotonic and potentially attracting or repelling kinks, with a saddle equilibrium separating the two.
Finally, we also explore the interplay of the two derivatives,
when both Caputo temporal and Riesz spatial derivatives
are involved.
\end{abstract}


\end{frontmatter}

\section{Introduction}

The study of physical, chemical and biological applications  of fractional
calculus has been a theme of growing interest over the
past few years. Relevant applications extend from
optical media~\cite{malomed,Longhi:15} to
epidemiological models~\cite{qureshi2020real},
and from biological systems~\cite{IonescuCNSNS2017}
to economics~\cite{ming2019application}.
By now, a diverse and broad range of corresponding
models has been summarized in books and associated reviews. Some of them
are more topical, as, e.g.,~\cite{cuevas}, focusing
on fractional models of dispersive waves, while others
are of more general interest, see e.g.,~\cite{podlubny1999fractional,Samko}.

The field of wave dynamics has been a fertile
platform for the investigation of such ideas,
as has been demonstrated by the introduction  of
the fractional Schr{\"o}dinger equation
in optics, through the work of~\cite{Longhi:15}.
Experimental realizations of related ideas, in the linear regime, have also appeared in the literature, see e.g.~\cite{malomed}.
A wide palette of analytical and numerical results, linear as well as nonlinear, continuum and also (long-range) discrete were recently reported in~\cite{cuevas}. Indeed, such wave phenomena hold
considerable promise for further experimental realizations,
including the nonlinear regime, given the extensive efforts of many scientists to engineer and control dispersion, spearheaded in
recent years by the works of~\cite{BlancoRedondoNC2016,RungeNP2020}; see
also~\cite{TamOL2019,TamPRA2020}. Very recently,
the first nonlinear waves experiment ---to the best
of our knowledge--- in a spatially
fractional dispersive setting was reported
in the work of~\cite{hoang2024observationfractionalevolutionnonlinear}.
{{Such an experiment in the realm of nonlinear optics
paves the way for future tailoring of dispersive properties in
optics, as well as in other fields and renders especially timely
the consideration of {\it prototypical} nonlinear, fractional
dispersive partial differential equation models, such
as fractional variants of the Korteweg-de Vries equation, the nonlinear Schr{\"o}dinger equation and the sine-Gordon equation
among others~\cite{cuevas}.

With that spirit and motivation in mind}}, in the present work, we build on recent studies of spatially fractional models of the $\phi^4$ type in~\cite{decker2024fractionalsolitonshomotopiccontinuation}
by examining the famous sine-Gordon equation~\cite{CKW14}
with fractional derivatives in both space and time.
While the relevant integer derivative models first appeared in the context of constant negative curvature surfaces~\cite{mclachlan},
the sine-Gordon equation has also often emerged in numerous other subfields of solid state physics, biophysics and beyond. For instance, in its discrete form it is a widespread model of the dynamics of crystal
dislocations~\cite{Frenkel1939}, while it has also been thoroughly
examined in the context of
Josephson junction arrays~\cite{Barone1982} and the dynamics of DNA models,
among other areas~\cite{Malomed2014}.

More precisely, we consider herein the modification of the
prototypical sine-Gordon model under the effect of spatial Riesz
derivatives, motivated in part by the possibility to engineer
dispersion~\cite{BlancoRedondoNC2016,RungeNP2020,hoang2024observationfractionalevolutionnonlinear}, as well as the introduction of temporal Caputo derivatives. The
latter is also motivated by the earlier work of two of the present authors~\cite{maciasbountis2022}, where suitable
discretizations for such models were first introduced.

After setting up the general framework of our model
in section 2, we proceed to examine three distinct scenarios:
In section 3, we vary the order of the temporal (Caputo) derivative,
while preserving the sine-Gordon (sG) right hand side.
{{Here, we see, for instance, how kinks slow down
in the presence of the Caputo temporal fractional derivatives
and appear to exhibit a dissipative effective motion.}}
In section 4, we return to the temporal second derivative and allow the spatial derivative order to change in the evaluation of the Riesz derivative.
{{An important finding in this setting is the
disparity between the case of Riesz exponents above the value
for the harmonic case (where there is a stationary kink-antikink
bound state with non-monotonic tails) and the exponent below
the harmonic, leading to a power law decay which is, however,
always monotonic in this limit.}}
Then, in section 5, we combine these variations. In each case, we consider first the dynamics
of the single kink (the prototypical coherent structure),
before exploring multi-kink or bound states thereof in the form of
kink-anti-kink or breather initial conditions.
{{In this setting, the above features when combined, give rise to intriguing kink-antikink
and breather dynamics, including, e.g., kink-antikink annihilation
or separation, and breathers that produce a kink-antikink pair that
may or may not decay at long evolution times.}}
Finally, in section 6 we summarize
our findings and discuss some directions for possible future
studies.

\section{Model Setup}

The general problem we study here concerns different types of solutions of the fractional sine-Gordon equation
\begin{eqnarray}
\frac{\partial^\beta u}{\partial t^\beta}=
\frac{\partial^\alpha u}{\partial x^\alpha} - \sin(u).
\label{eq:sinegordon}
\end{eqnarray}
For the integrable case $\alpha=\beta=2$, the solutions of Eq.~(\ref{eq:sinegordon}) are well-known~\cite{CKW14}.
To appreciate the effect of fractional derivatives, we
consider the following cases of interest:
\begin{enumerate}
    \item Caputo derivative of order $\beta \neq 2$ in time,
    but second derivative in space.
    \item Riesz derivative of order $\alpha \neq 2$ in space,
    but second derivative in time.
    \item The general case, where a
    Caputo derivative of order $\beta \neq 2$ in time is combined
    with a Riesz derivative of order $\alpha \neq 2$ in space.
\end{enumerate}
Our aim in all cases is to explore the dynamics of the single
kink, as well as the kink-anti-kink pair (and related structures
of particular relevance to the sine-Gordon continuum model~\cite{birnir})  in the presence of fractional derivatives.
{{Our motivation here is analogous to
the recent development of~\cite{hoang2024observationfractionalevolutionnonlinear} within
nonlinear optics. If in any of the diverse settings mentioned
in the Introduction (see e.g. an over-arching summary the book
of~\cite{CKW14}), the possibility to experimentally manipulate/engineer
dispersion similarly arises, then Eq.~(\ref{eq:sinegordon}) emerges
as the canonical fractional derivative extension thereof.}}

We consider Caputo derivatives, instead of, e.g., those of Riemann-Liouville type, as in the former case the initial conditions  are immediately retrieved from the integer derivative case, see e.g.~\cite{podlubny1999fractional,Samko} for relevant details.
{{Indeed, one can then specify $u(0,x)$ and $\partial_t u(0,x)$,
as is, in general, natural for initial value problems such as the
ones specified herein.}}
The algorithm used for numerical integration is \texttt{fde\_pi12\_pc}, which can be found in \cite{Garrappa}.

\section{Caputo Derivative in Time, Laplacian in Space}
\label{justcaputo}

We start with case 1 above: preserving the
spatial Laplacian and only modifying the time derivative
in the form of a variable Caputo exponent.

\subsection{Kink dynamics}

First, we examine the most prototypical waveform solution of the model,
namely the single kink. We consider as our Riesz exponent $\alpha=2$ and values of the  Caputo exponent $\beta<2$. We also comment on the case
with $\beta>2$ later on. As initial condition for Eq.~\eqref{eq:sinegordon}, we consider

\begin{equation}
     ((u(x,0),\dot{u}(x,0))=(u_0,\dot{u}_0)=(4 \arctan{e^{\gamma_0 x}}, -2\gamma_0  v_0 \sech(\gamma_0 x)),
    \label{kink1}
\end{equation}
where the overdot stands for the partial derivative in time and
\begin{equation}
    \gamma_0=\frac{1}{\sqrt{1-v_0^2}}
    \label{kink11}
\end{equation}
represents the Lorentz contraction factor~\cite{CKW14}.

\begin{figure}[h]
\begin{tabular}{cc}
\includegraphics[width=.45\textwidth]{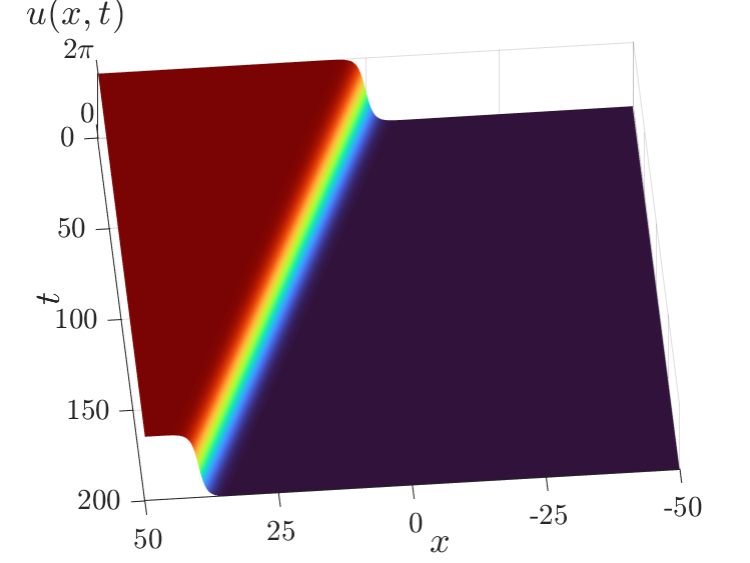} &
\includegraphics[width=.45\textwidth]{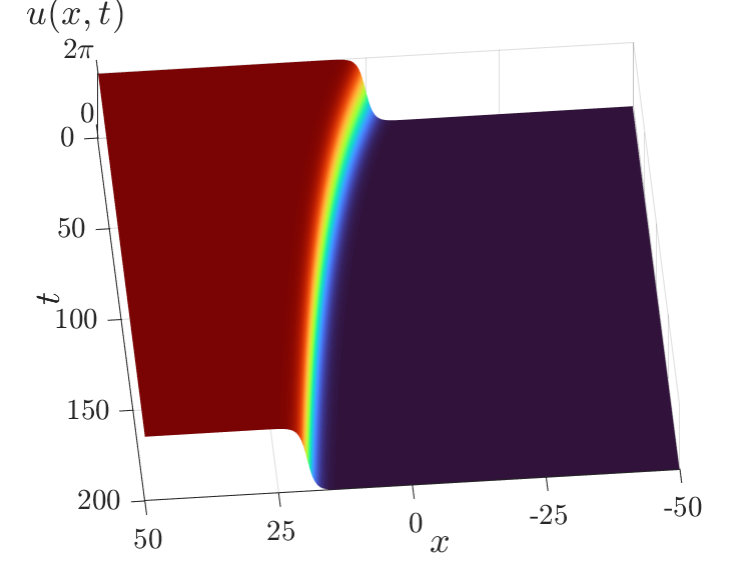} \\
\end{tabular}
{\caption{Time evolution of the kink with initial velocity $v_0=0.2$ for the ordinary Laplacian ($\alpha=2$). Panels show $u(x,t)$ and compare the Hamiltonian constant-velocity dynamics of $\beta=2$ (left) and the decelerating dynamics of the Caputo-fractional case of $\beta=1.9$ (right). Notice how the former panel showcases a traveling
kink, while for the latter the kink trajectory starts ``bending'',
reflecting the kinks temporal slowdown.
\label{fig:example}}}
\end{figure}

\begin{figure}[h]
\includegraphics[clip,width=0.5\textwidth,trim=0cm 0cm 0cm 0cm]{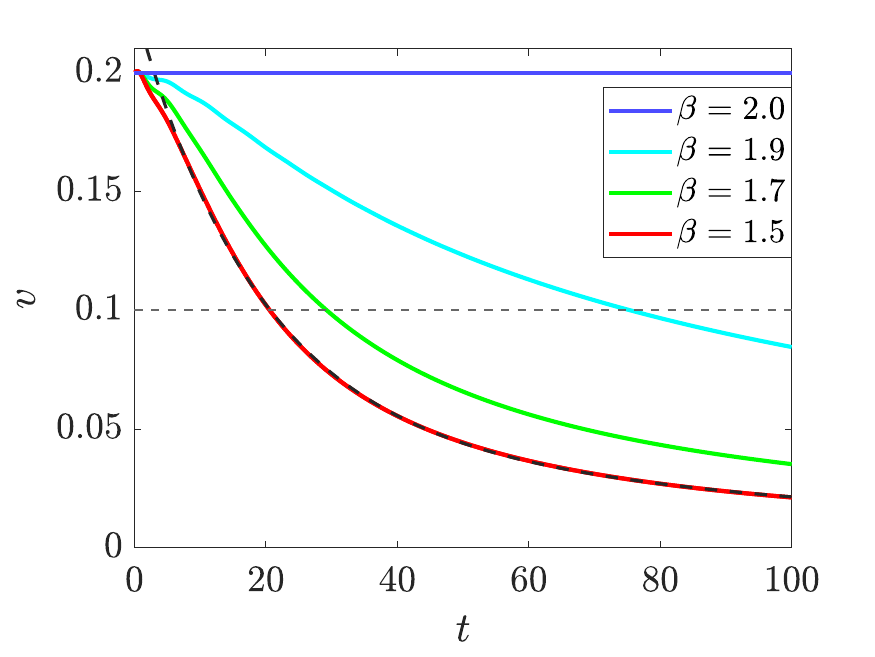}
\includegraphics[clip,width=0.5\textwidth,trim=0cm 0cm 0cm 0cm]{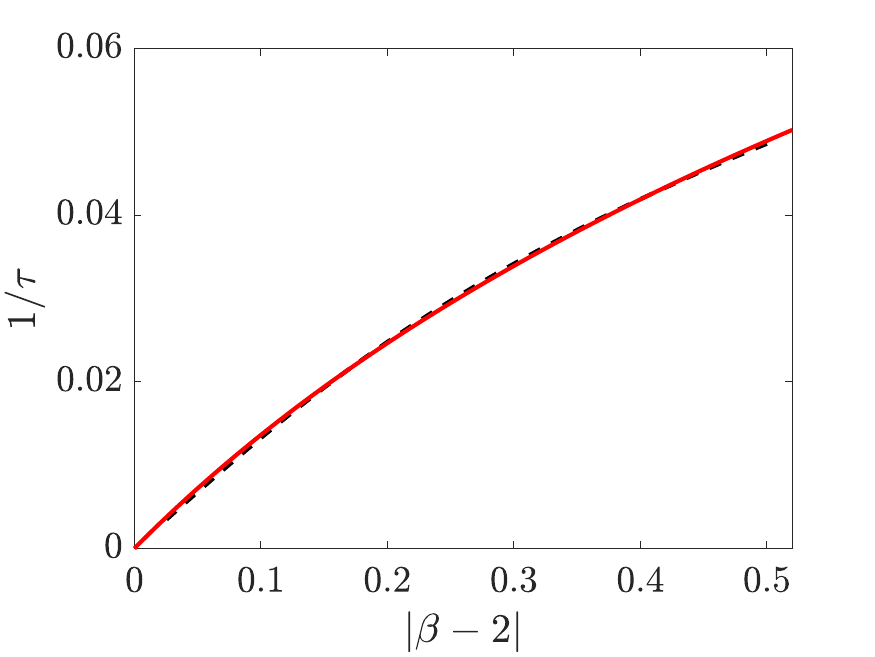}
\caption{Left panel: Kink velocity decay with the order of the Caputo derivative. For $\beta=2$, the kink propagates at a constant velocity. The fractionality acts as form of damping, exponentially decreasing the velocity. For $\beta=1.5$, the exponential fit is represented by a dashed black line. We define $\tau$ as the time for which the velocity halves ($v=0.1$, also marked here in gray) for each $\beta$ value.
Right panel: Inverse of the characteristic time with respect to the order of the Caputo derivative. As $\beta$ approaches 2, the characteristic time tends to infinity (i.e., $1/\tau$ tends to zero), since the kink motion is undamped in that limit. The dotted black line depicts the numerical results and the red solid line is a logarithmic fit.
\label{fig:caputo_1}}
\end{figure}

For $\beta=2$, the kink propagates at a constant speed $v=v_0$, which we fix to 0.2, as expected from the corresponding exact solution. However, when fractionality is introduced in the equation, and we consider derivative orders $\beta<2$, the kink velocity \textit{decays in time}. {This is shown in Fig.~\ref{fig:example}, where spatio-temporal plots of $u(x,t)$ for $\beta=2$ and $\beta=1.9$ are compared. In addition, left panel of Fig.~\ref{fig:caputo_1}} depicts the kink velocity evolution as a function of time. Here, we can clearly observe the dissipative effect of the Caputo derivative. The kink naturally slows down and {the velocity eventually tends to zero}. Indeed, similar features were first reported for breathers in the time-fractional sine-Gordon equation \cite{maciasbountis2022,BOUNTIS2024100807}, where the oscillations were first shown to be damped for values of $\beta<2$. It is this parameter setting that we also explore
herein.

Remarkably, the velocity of the kink decays exponentially over time after an initial short (transient) interval. This is clearly seen in
the left panel of Fig.~\ref{fig:caputo_1}, where the exponential fit is depicted as a black dashed line for $\beta=1.5$ in excellent agreement with the PDE results. We define by $\tau$ the characteristic time at which the velocity halves, and mark it in the figure with an horizontal gray dashed line. In the right panel of Fig.~\ref{fig:caputo_1}, we plot the variation of the inverse of characteristic time with respect to the order of the Caputo derivative.  As $\beta$ approaches the value $2$, $\tau$ tends to infinity and the inverse tends to zero, as one would expect, based on the absence
of dissipation in that limit. In the same panel, we included a logarithmic fit of the form: $1/\tau=a \cdot \log(b\cdot |\beta-2|+1)$, with $a=0.0581$, $b=2.6410$ and $r=0.9998$.

\begin{figure}[h]
\begin{tabular}{cc}
\includegraphics[width=.45\textwidth]{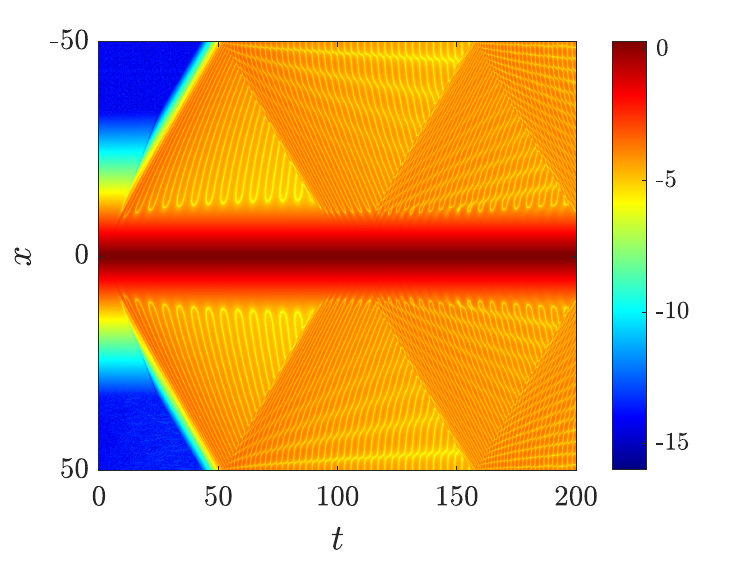} &
\includegraphics[width=.45\textwidth]{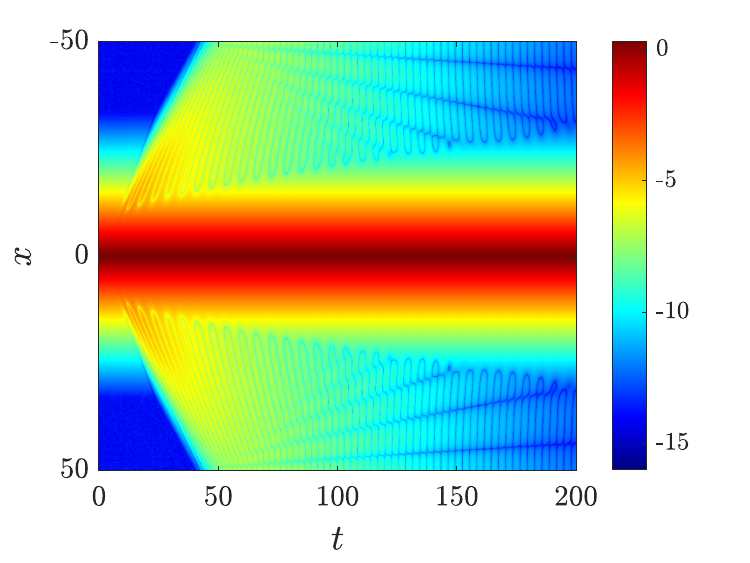} \\
\end{tabular}
\caption{Time evolution of the pulse-like derivative of a stationary kink with weak (numerical scheme induced) radiation on top of an exact stationary kink
solution, when fixed ends boundary conditions and ordinary Laplacian ($\alpha=2$) are considered. Panels show $\log_{10} |\partial_x u(x,t)|$ and distinguish between the radiation-reflecting case of $\beta=2$ (left) and the ``radiation-removing'' case of $\beta=1.9$ (right). Thus, radiation is ``distilled" (eliminated) in the Caputo-fractional case of $\beta=1.9$, while it hits the boundary and returns in the Hamiltonian ($\beta=2$) case.
\label{set_2}}
\end{figure}

It is interesting to complement the above results
with the case of a kink that is stationary, in part
 because this is the ultimate fate of the structure
in the model bearing a Caputo derivative, but also to
compare/contrast the relevant phenomenology with that
of the standard (conservative) sine-Gordon case. Thus, when initializing
Eq.~(\ref{eq:sinegordon}) for vanishing velocity $v_0=0$, one observes
the left panel of Fig.~\ref{set_2} for $\beta=2$, and
the right panel for $\beta=1.9$.
In the left panel, we show in logarithmic
scale the pulse-like derivative of the kink and we observe that
the approximation induced numerically by the local truncation error
leads to the propagation of
radiation wakes that move towards the boundary and eventually
get reflected from it, returning to the kink
and impinging upon it. The orange dominant color in these
radiation waves reflects
this discretization-induced error, for
our spatial discretization of spacing of O$(10^{-1})$.
Importantly, the radiation never gets ``washed away'' given
the Hamiltonian nature of the model.
It is relevant to note here that, in this stationary
case, had we presented the 3d profile of $u=u(x,t)$, the left
and right panels would be indiscernible as the $0-2 \pi$ scale
of such panels would not permit to distinguish their much
smaller scale differences.

On the contrary, the right panel of the figure shows quite a
different picture for the case of a Caputo
derivative with $\beta=1.9$. What we see is that the
color associated with the radiation wakes drastically changes
from orange to yellow and gradually to green and blue, reflecting
the dissipative nature of the corresponding derivative
and essentially the ``annihilation'' of the relevant wakes, or,
otherwise stated, the ``distilling'' of the kink into the
appropriate stationary state of the model.
That is to say, these computations, both at the
dynamical level of the slowing down moving kink, but also
at the stationary level of the static kink reflect the
dissipative nature of the  effect of the Caputo temporal
derivative.

\FloatBarrier

\subsection{Kink-anti-kink dynamics}

Let us now consider the dynamics of head-on collisions of kink-anti-kink pairs in the Caputo time derivative case with $\beta<2$. To this end, we assume an initial condition of the form

\begin{equation}
    (u_0,\dot{u}_0)=(4 [\arctan{e^{\gamma_0 (x+\delta)}}-\arctan{e^{\gamma_0 (x-\delta)}}], -2\gamma_0  v_0 [\sech(\gamma_0 (x+\delta))+\sech(\gamma_0 (x-\delta))],
    \label{kink2}
\end{equation}

If $\beta$ is slightly smaller than 2, we observe that, if $v_0$ is above a certain threshold, the dynamics is similar to the $\beta=2$ case, namely the collision is quasi-elastic, followed by an inversion of the kink-anti-kink roles, {and a subsequent deceleration of the kink/anti-kink similar to that observed previously for the individual kink}. Below this threshold velocity $v_0$, the kink-anti-kink pair collapses to a breather-like structure that decays in time in a form similar to what was observed in \cite{BOUNTIS2024100807}.
For a value of $\beta=1.95$ the threshold $v_0$ is approximately $0.590$.

\begin{figure}[h]
\begin{tabular}{cc}
\includegraphics[width=.45\textwidth]{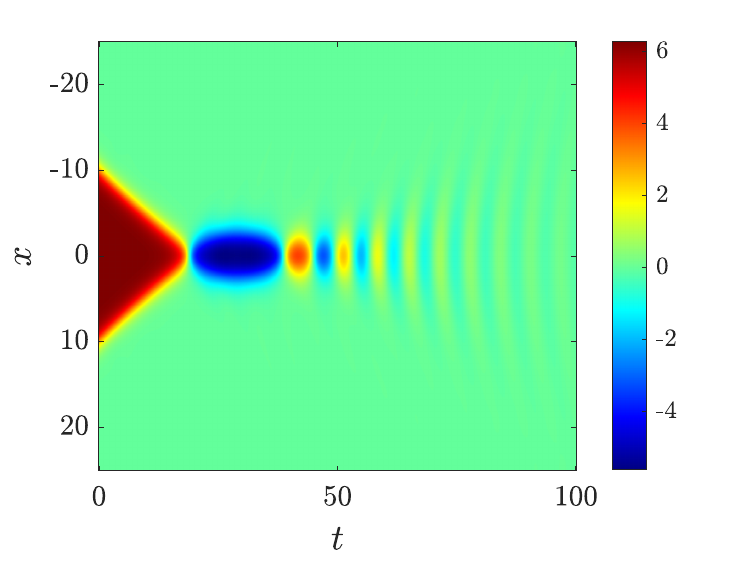} &
\includegraphics[width=.45\textwidth]{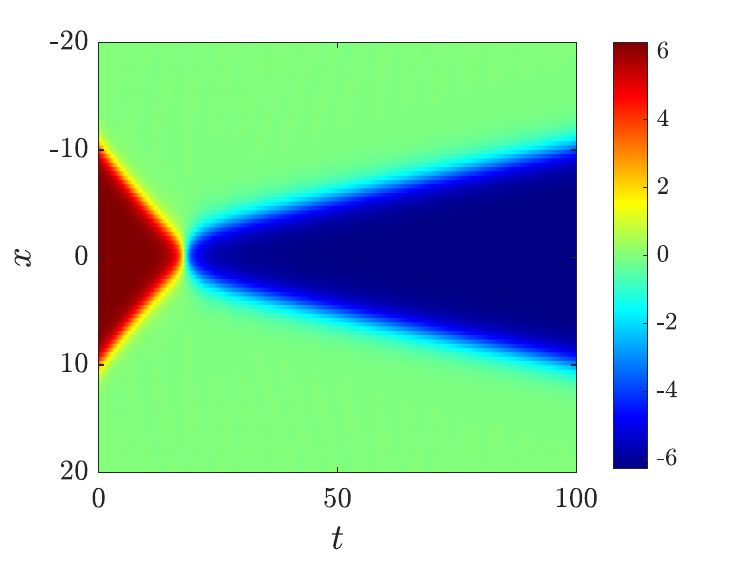} \\
\end{tabular}
\caption{{The evolution of a kink-anti-kink pair initially separated by a distance $2\delta=20$, launched with an initial velocity $v_0$, for $\beta=1.95$ and $\alpha=2$. The left panel shows the generation of a subsequently vanishing breather when $v_0=0.55$ whereas the right panel corresponds to a quasi-elastic collision occurring when $v_0=0.6$. The critical value separating the two regimes is $v_0\approx0.590$.}}
\label{fig:collision}
\end{figure}

Figure~\ref{fig:collision} shows, for $\beta=1.95$, representative examples of these behaviors. More specifically, these simulations suggest that the damped  kink-anti-kink interaction dynamics is still subject  (as in the standard sG model) to an attractive interaction
potential well with finite depth~\cite{CARRETEROGONZALEZ2022106123}.

Contrary to the standard integrable sG case, however, where the dynamics inside this well is conservative, here the dynamics is dissipative, as shown clearly on the
{left} panel of Fig.~\ref{fig:collision}. Thus, in order
to escape, the kinks need to overcome both each other's
attraction and the reduction of their speed due to dissipation.
Understanding the energetics and precise dynamics of
the corresponding waveforms as a function of $\beta$ is certainly
an intriguing question that needs further exploration.

When $\beta$ decreases, the critical $v_0$ increases and, if $\beta$ is far enough from $2$, this critical value is very close to 1 and, {at the same time the kinks, progressively widen before they interact}. It is worth mentioning that the critical $v_0$ also depends on the initial separation $2\delta$, so that $v_0$ increases when $\delta$ increases. This is clearly due to the individual dissipative effect of the Caputo derivative on each kink prior to their interaction.

\FloatBarrier

\section{Riesz Derivative in space, {conservative ($\beta=2$)} dynamics in time}

Having explored the case of varying the order of the time derivative, we now turn to the examination
of what happens when the order of the space derivative is varied. In particular, we examine the spatial Riesz derivative
order $\alpha$ and the impact of its variation, while we keep the temporal derivative at $\beta=2$.

\subsection{Kink dynamics}

\begin{figure}[h]
\begin{tabular}{cc}
\includegraphics[width=.45\textwidth]{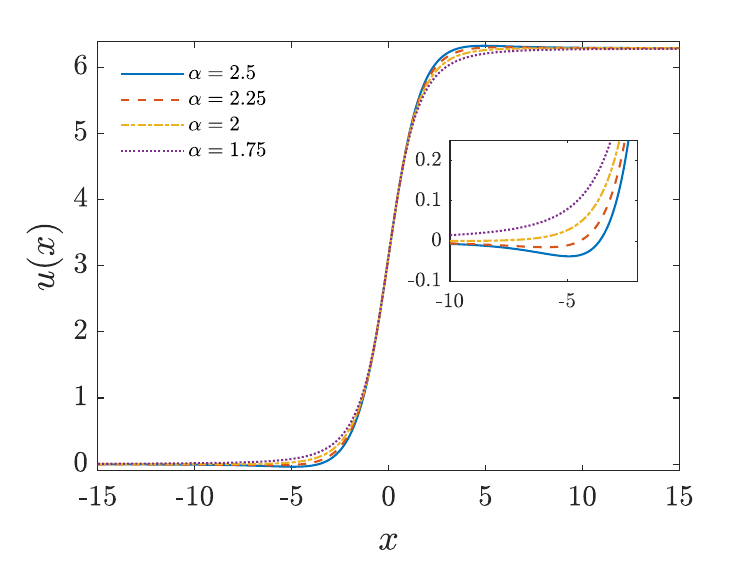} &
\includegraphics[width=.45\textwidth]{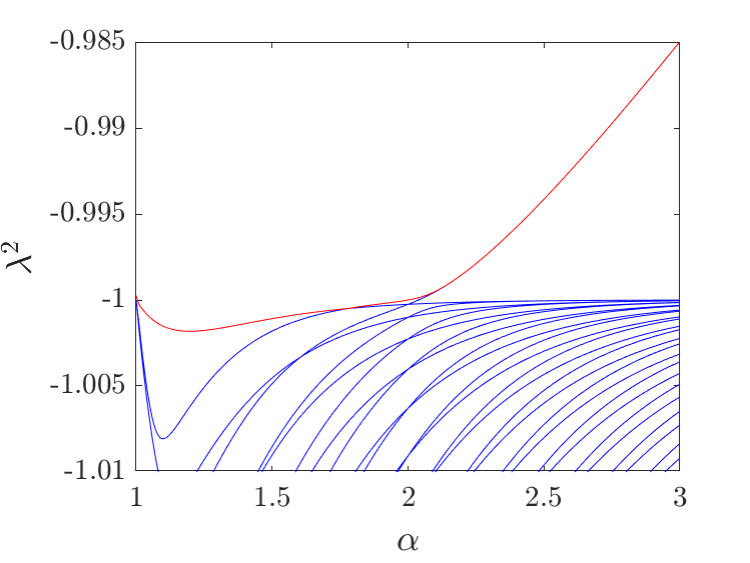} \\
\end{tabular}
\caption{Single kink, Riesz Derivative of fractional order
$\alpha$ in space, with $\beta=2$ in time. The left panel shows
the stationary kink for 4 different values of the spatial derivative
order $\alpha$, namely $\alpha=2.5$, $\alpha=2.25$
$\alpha=2$ and $\alpha=1.75$.
Observe the {\it non-monotonic} nature of the kink for
$\alpha>2$, while the structure becomes monotonic for $\alpha \leq 2$.
The right panel shows the (squares of the) lowest non-zero eigenvalues of the kink
spectrum as a function of $\alpha$. The eigenvalue positioned
at the band edge $\lambda = \pm i$ in the integrable case of
$\alpha=2$, shown here in red, moves inside the continuous spectrum for $\alpha<2$,
while it becomes an ``internal mode'' of the kink for $\alpha>2$. The
kink is spectrally
stable for all considered values of $\alpha$.
\label{set_4}}
\end{figure}

Once again, our first consideration involves the single
kink dynamics. Here, we perform a study reminiscent
of the recent $\phi^4$ model exploration carried out in ~\cite{decker2024fractionalsolitonshomotopiccontinuation}.
In particular, we point out that a single kink has a
\textit{monotonic} spatial profile for $\alpha<2$, as shown
on the left panel of Fig.~\ref{set_4}, while this
changes to a non-monotonic state involving a single
crossing of the uniform steady states (of $u=0$ and
$u=2 \pi$) on each side when $\alpha>2$. This non-monotonicity
will be seen below to be connected to the
existence of a potential kink-anti-kink (or K-AK) equilibrium in this model.
The inset of the figure more clearly highlights the
relevant spatial behavior.

On the other hand, in this setting we not only explored
the existence problem, identifying the relevant stationary
kink; we also examined the corresponding linearization
problem associated with spectral stability. The relevant
eigenvalue problem is based on the ansatz
$u(x,t)=u_0(x) + \epsilon e^{\lambda t} w(x)$, where
$u_0(x)$ is the spatially stationary configuration
(in the context of this subsection the single kink) and
leads to the form:
\begin{eqnarray}
\lambda^2 w=
\frac{\partial^\alpha w}{\partial x^\alpha} - \cos(u_0) w.
\label{linearize}
\end{eqnarray}
The results of the spectral analysis are shown on the
right panel of Fig.~\ref{set_4}. Here, we see that the
mode (in red) sitting at the band edge in the form of
a resonance for the standard integrable sine-Gordon
case bifurcates from there for $\alpha>2$ and, indeed,
becomes an internal breathing mode of the kink coherent
structure.
Such internal modes have been of considerable interest
in the context of models such as the perturbed
sG equation~\cite{IM}. On the other hand, for the case of
$\alpha<2$, the kink follows an opposite trajectory
moving inside of the continuous spectrum.
In either case here, this does not affect the
spectral stability of the kink, but may affect
the outcome of kink collisional interactions, as is
well known to occur in the presence of relevant internal
modes~\cite{CAMPBELL19831}. Such a topic is outside
the scope of the present work and may be of interest
in a future study.

\FloatBarrier
\subsection{Kink-anti-kink and breather dynamics}

\begin{figure}[h]
\begin{tabular}{cc}
\includegraphics[width=.45\textwidth]{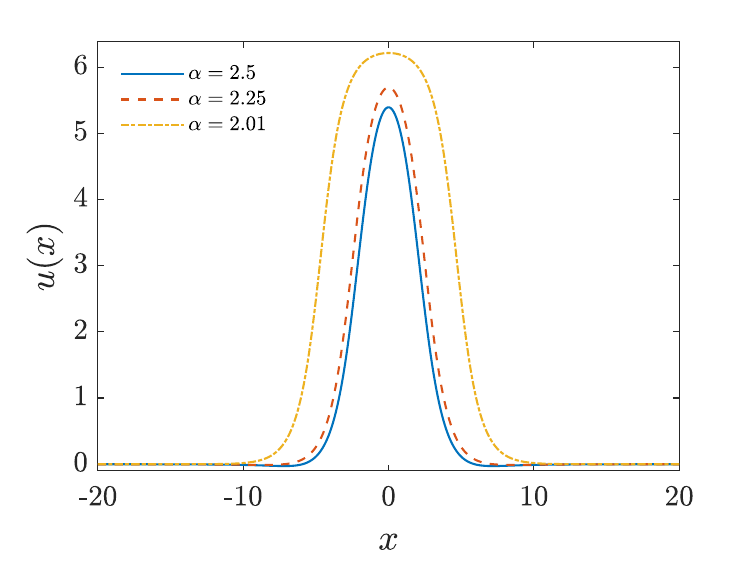} &
\includegraphics[width=.45\textwidth]{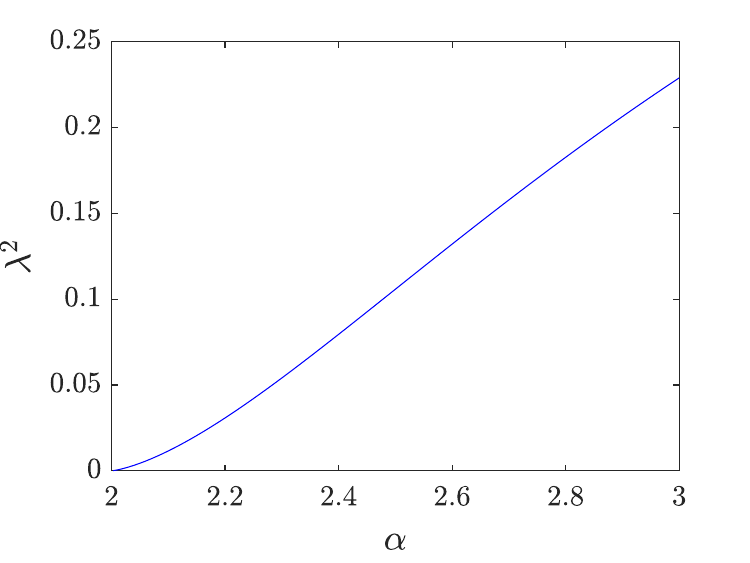} \\
\end{tabular}
\caption{Left panel: Kink-anti-kink stationary configuration
for a Riesz Derivative of fractional order
$\alpha$ in space.
The kinks are pinned symmetrically about $x=0$ and separated by a distance that depends on $\alpha$.
Indeed, a bound
state exists $\forall \alpha>2$. Three different values of $\alpha$ are considered, namely
$\alpha=2.5$, $\alpha=2.25$ and $\alpha=2.01$.
It can be seen that the distance between the kink and the anti-kink
of the bound state {\it diverges} as the fractional order $\alpha$
approaches the limit $\alpha \rightarrow 2^+$. Right panel: The eigenvalue pertaining to this kink-anti-kink
bound state. We can see that this state is
{\it spectrally unstable} $\forall \alpha >2$, with the relevant
eigenvalue becoming larger, the larger the deviation of $\alpha$ is
from its threshold at $\alpha=2$.
\label{set_5}}
\end{figure}

The examination of K-AK pairs in the setting of the
Riesz spatial derivative is quite interesting in its
own right, as manifested in Fig.~\ref{set_5}.
Here, we can see that for all values of $\alpha>2$, a
K-AK bound state pair exists. Indeed, this pair bifurcates
from $\infty$, as the separation between K and AK
diverges for $\alpha \rightarrow 2^+$. Such an equilibrium
is closely intertwined with the non-monotonic nature of the
tail and the existence of the zero crossing within such a tail.
It is well-known for some time~\cite{manton1979effective} that the nature of the solitary wave tail is responsible for the
force between the solitary waves, as the latter is
mediated through the tails. Accordingly, our observation
in the previous subsection about the existence of a zero
crossing of the kink tail leads to the emergence of
a zero-crossing of the force between a kink and an anti-kink.
This, in turn, is responsible for the emergence of a K-AK bound state, as illustrated for different values of the Riesz exponent
$\alpha$ in Fig.~\ref{set_5}.

When analyzing the stability of this configuration,
as shown in the right panel of Fig.~\ref{set_5}, we find
that the relevant state is {\it generically} unstable.
The corresponding unstable eigenmode has an eigenvalue which
tends to $0$, as the K-AK separation tends to $\infty$,
restoring the stability of isolated kinks in the
limit of $\alpha \rightarrow 2^+$. At the same time,
exploring the eigenvector associated with the instability
we see that it is related to the K-AK out-of-phase
motion. At the same time, the very fact that this is
an unstable fixed point of the PDE strongly suggests that this
K-AK stationary configuration represents an energetic
maximum (i.e., a saddle point) in this Hamiltonian system.

\begin{figure}[h]
\begin{tabular}{cc}
\multicolumn{2}{c}{\includegraphics[width=.45\textwidth]{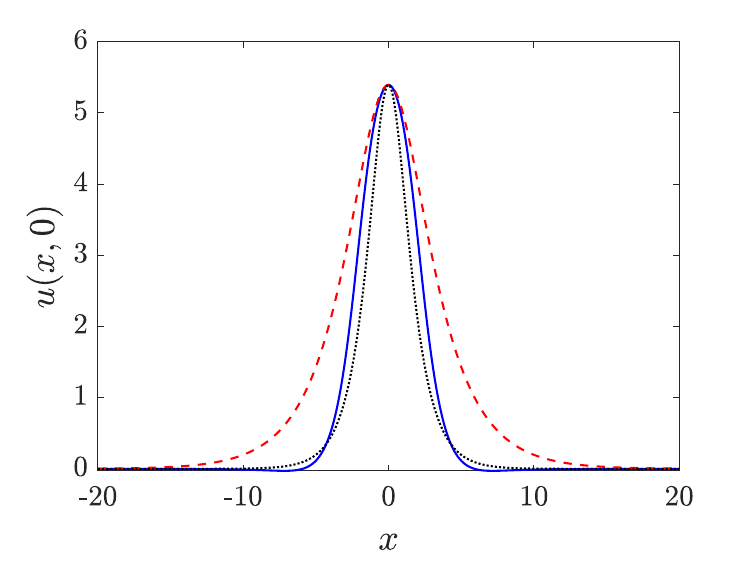}} \\
\includegraphics[width=.45\textwidth]{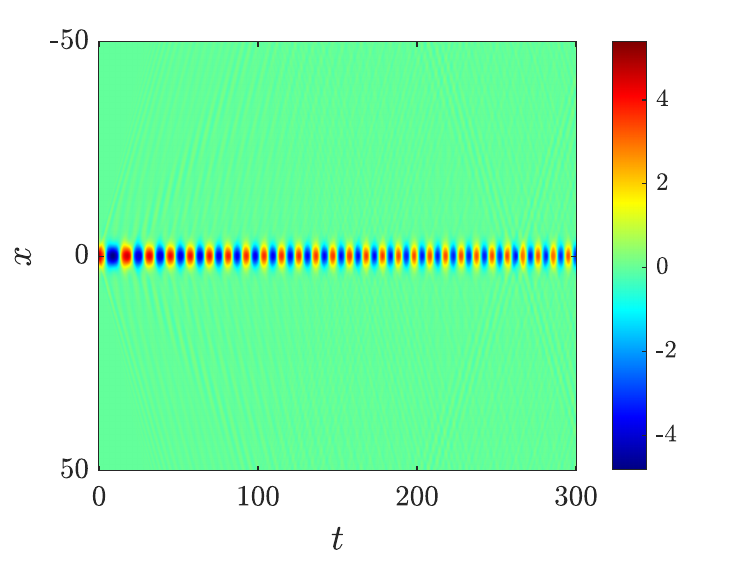} &
\includegraphics[width=.45\textwidth]{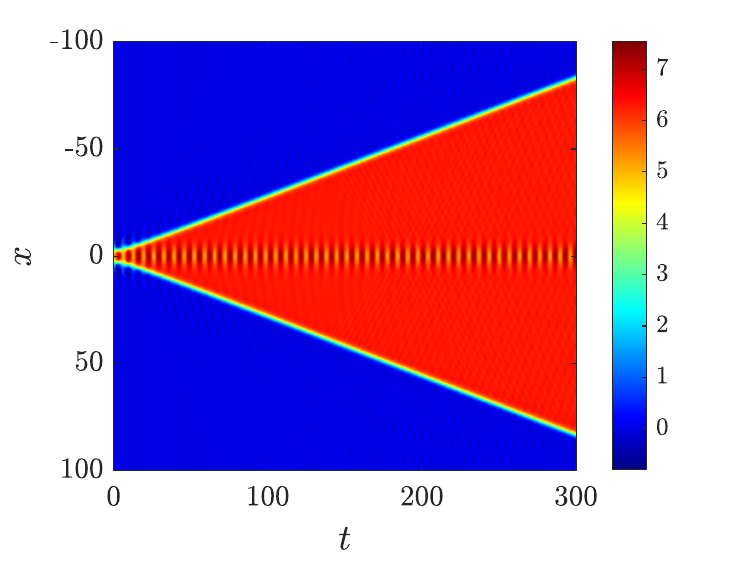} \\
\end{tabular}
\caption{Dynamics of the kink-anti-kink bound state for fractional
order $\alpha=2.5$ (typical of $\alpha>2$).
In the top panel the blue solid curve
represents the equilibrium configuration for $\alpha=2.5$. The
red dashed line corresponds to the initial condition
$u(x,0)=A\,{\rm sech}(0.4x)$ (with $A\approx5.3719$ being the maximum amplitude of the K-AK bound state), reflecting a ``wider'' configuration,
while the black dash-dotted line corresponds to the initial condition
$u(x,0)=A\,{\rm sech}(0.8x)$, pertaining to a ``narrower'' initial
condition. The latter can be seen in the bottom left panel to lead the kink and the anti-kink attracting each other and
colliding indefinitely forming a breathing state (while emitting radiation). The widening K-AK initial state, on the other hand,
lying ``outside'' of the energy maximum, yields a space-time evolution
associated with an indefinite separation of the kink and the anti-kink (see bottom right panel).
\label{set_6}}
\end{figure}

In turn, this has implications for the dynamics shown in Fig.~\ref{set_6}. Here, kinks with separation lower than a critical value are expected
to collapse into each other due to their pairwise attraction.
On the other hand, kinks with separations larger than this
value are expected to separate indefinitely
due to their repulsion (with the force changing from attractive
to repulsive at this critical point). We select two configurations (for a fixed amplitude coincident with the amplitude of the equilibrium K-AK state), which are parametrized by the inverse width of the
configuration. The red dashed line represents a configuration
where K and AK have a separation larger that the (equilibrium)
critical one, while the black dash-dotted line pertains to a smaller separation. Indeed, the latter are shown at the bottom of the left panel to be indefinitely bound to each other, while the former
configuration with the structures being past the
energy landscape's saddle point, can be seen to lead
to indefinite separation of the kinks, in line with our theoretical expectation.

Lastly, in the context of the Riesz derivative variation,
we explore the dynamics of breather configurations
(as a variant of the K-AK structure).
One of the consequences that can be derived from the
bottom left panel of Fig.~\ref{set_6} is, indeed, that breathers exist for the model with $\beta=2$ and $\alpha\neq2$. Although some numerical experiments were performed in \cite{Alfimov}, they were not directly computed as periodic orbits.
Here, we perform such a computation by using Fourier-space techniques similar to those developed in \cite{Martina}.

\begin{figure}[tbp]
\begin{tabular}{cc}
\includegraphics[width=.45\textwidth]{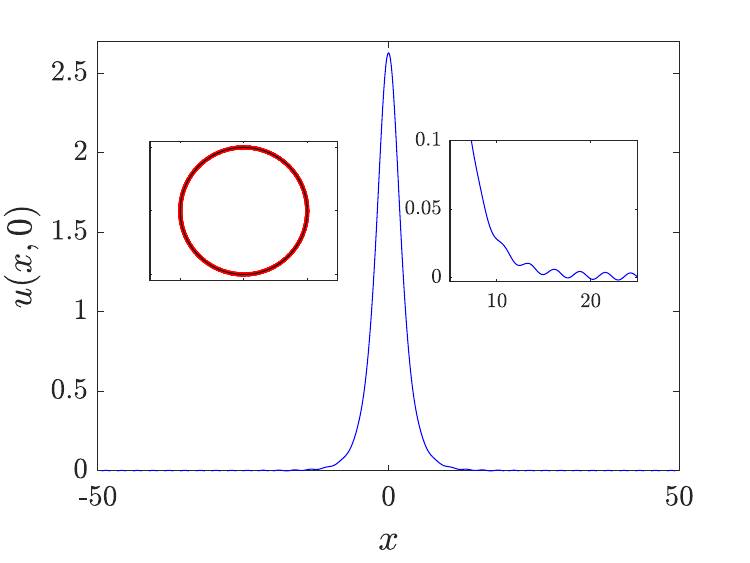} &
\includegraphics[width=.45\textwidth]{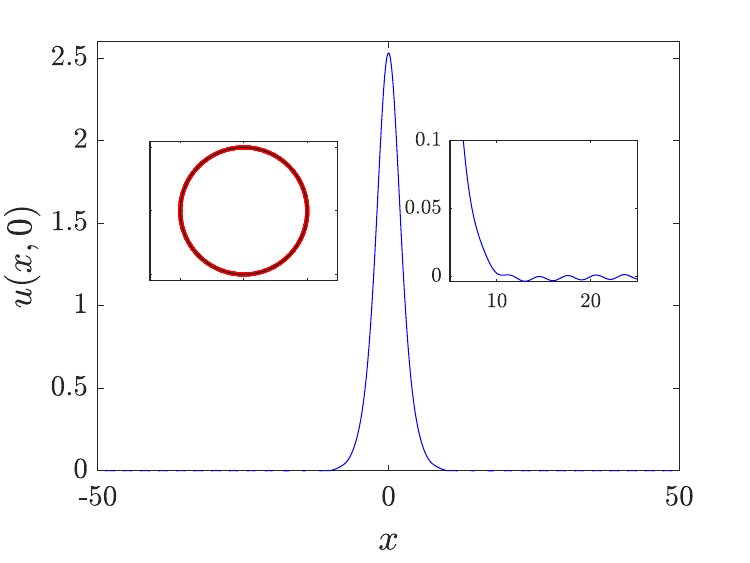} \\
\end{tabular}
\caption{Breathers with $\beta=2$ and $\alpha=1.85$ (left) and $\beta=2.15$ (right).
Their frequency is $\omega=0.8$ and their domain is $[-50,50]$. Insets show a zoom on the breather wings (right) and the Floquet multipliers spectrum (left).}
\label{fig:breathers}
\end{figure}

In particular, we show in Fig.~\ref{fig:breathers} the profile of breathers in two cases, one with $\alpha<2$ and another with $\alpha>2$,
for the frequency $\omega=0.8$. In this setting,
the second and subsequent harmonics of the frequency lie within the
continuous spectrum.
As can be deduced from the figure (and the insets,
featuring the breather tails, as well as their stability), there are no qualitative differences between the two cases. In addition, they feature small wings, as expected because of the breaking of
the integrability of the sine-Gordon equation and the resonance
of the breather's n-th harmonic frequency with the continuous spectrum, when the Laplacian is fractional.

As a consequence, the breather is actually a so-called \textit{nanopteron}, as typically occurs in continuous models with finite domains~\cite{CKW14}. Because of the resonances with the linear modes, such models are connected with complex bifurcation diagrams; see~\cite{caputo2} for a relevant example.
As a consequence, in the examples shown in Fig.~\ref{fig:breathers} we have selected a couple of cases where the breather appears to be stable in the finite domain computation provided herein.
It is relevant to mention that, in order to assess the
spectral stability of the breather, standard Floquet multiplier
methods have been employed, based on the eigenvalues of
the monodromy matrix, as discussed, e.g., in~\cite{FLACH20081}.

\FloatBarrier
\section{Caputo derivative of order $\beta \neq 2$ and Riesz derivative of order $\alpha \neq 2$ }

Before discussing our results regarding the dynamics of kinks in this case, it is relevant to remark that, as demonstrated in the Appendix, the stability spectrum of stationary kinks for any value of $\beta$ is exactly the same as for $\beta=2$.
The relevant findings indeed apply to the results
of Section~\ref{justcaputo}, yet we mention this
feature here, as it is true for all values of
$\alpha$ as well. This suggests the intriguing feature
that for $\beta<2$, while the
kink has a spectrum lying on the imaginary axis
(i.e., with $\lambda^2<0$), the corresponding
temporal eigenfunction is responsible for its
decay.

\subsection{Kink dynamics}

Similarly to the case of $\alpha=2$, we consider here the dynamics of kinks with an initial velocity $v_0$ when $\beta<2$ is varied. To this aim, we take as initial condition a perturbed static kink $u_\mathrm{kink}(x)$ of the form:
\begin{equation}
    (u_0,\dot{u}_0)=(u_\mathrm{kink}(x), v_0 \partial_x u_\mathrm{kink}(x)),
    \label{ansatz}
\end{equation}
This choice of initial conditions is motivated by the difficulty of applying a Lorentz boost to the numerical function $u_\mathrm{kink}(x)$ (for values of
$\beta \neq 2$).

\begin{figure}[h]
\centering
\includegraphics[clip,width=0.45\textwidth,trim=0cm 0cm 0cm 0cm]{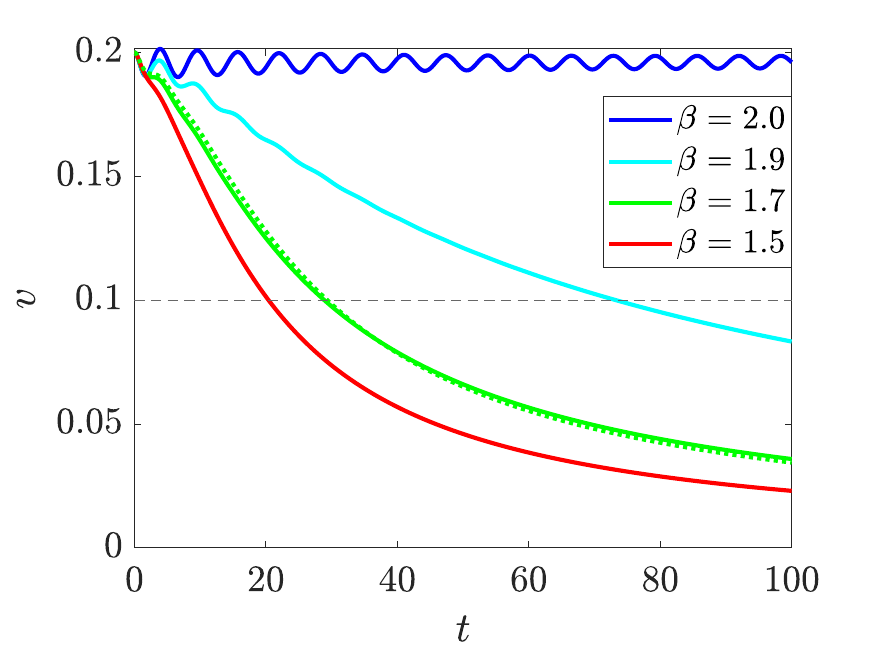}
\caption{Kink velocity decay with time for different values of the order of the Caputo derivative for $\alpha=1.7$. We have also included a dotted line for the case $\beta=1.7$ with a different value of $\alpha$, here $\alpha=2.5$, for comparison. It can be seen that the value of $\alpha$ barely affects the velocity decay. The initial velocity in all cases is $v_0=2$.
\label{fig:riesz_caputo_kink}}
\end{figure}

As can be seen in Fig.~\ref{fig:riesz_caputo_kink}, the dynamics of the kink is almost independent of the value of $\alpha$ and is predominantly dictated by the decay imposed by
the presence of the Caputo derivative.
Notice, for example, the decay for $\beta=1.7$, for which we present here for the cases of $\alpha=1.7$ (solid green line) and $\alpha=2.5$ (dotted green line). The curves are nearly identical.
Consequently, the dynamics for $\alpha\neq2$ is quite
similar as $\alpha=2$. The only remarkable difference is the existence of oscillations in the velocity for $\beta=2$ which is
somewhat artificial and is caused by the fact that
Eq.~(\ref{ansatz}) is not consonant with the Lorentz boost
which can be applied (cf. Eqs.~(\ref{kink1})-(\ref{kink11}))
in the latter case, giving rise to a traveling solution.

\subsection{Kink-anti-kink dynamics}

{In this section, we consider the Riesz derivative and Caputo derivative combined to further study the K-AK pair. First of all, we have performed a study similar to the case with $\alpha=2$, monitoring the dynamics of collisions between kinks and anti-kinks separated a distance $2\delta$. To this end, we take an initial condition of the form}

\begin{equation}
    (u_0,\dot{u}_0)=(u_\mathrm{kink}(x+\delta)+(2\pi-u_\mathrm{kink}(x-\delta)), v_0 [\partial_x u_\mathrm{kink}(x+\delta))+\partial_x u_\mathrm{kink}(x-\delta))]),
\end{equation}
with $u_{\mathrm{kink}}(x)$ being the stationary kink solution for such a value of $\alpha$. The observed evolution is qualitatively similar to that of Fig.~\ref{fig:collision}; for that reason, no
additional figures have been included. It must be noticed that, as expected, the critical initial velocity depends on both $\alpha$ and $\beta$ so that, for fixed $\beta$, the critical $v_0$ increases when $\alpha$ decreases.

\begin{figure}[h]
\begin{tabular}{cc}
\includegraphics[width=.45\textwidth]{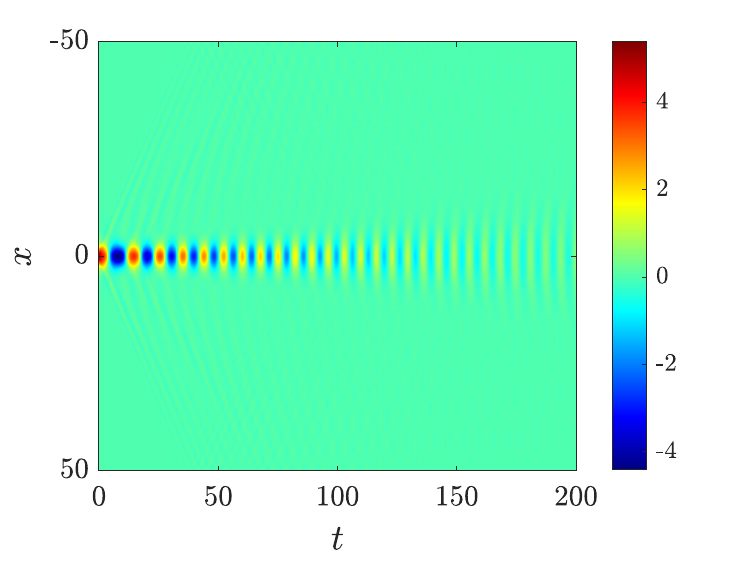} &
\includegraphics[width=.45\textwidth]{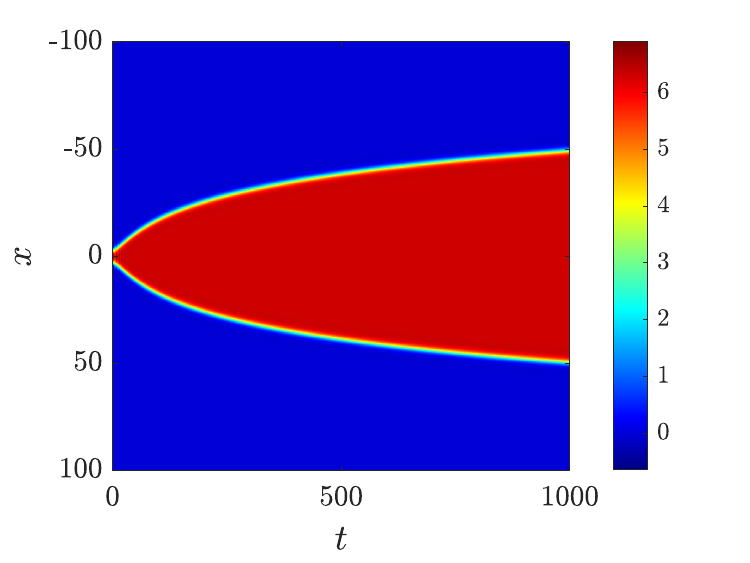} \\
\end{tabular}
\caption{{Dynamics of the kink-anti-kink bound state for fractional
order $\alpha=2.5$ and $\beta<2$ in a similar fashion to Fig.~\ref{set_6}. The left panel shows the outcome for $\beta=1.99$ and an initial condition
$u(x,0)=A\,{\rm sech}(0.8x)$ (with $A\approx5.3719$), resulting in a breather whose oscillation decays
over time. The right panel corresponds to $\beta=1.9$ and a ``wider'' initial condition
$u(x,0)=A\,{\rm sech}(0.4x)$, leading to a kink and anti-kink whose separation increases in time but in a decelerated fashion.}
\label{set_7}}
\end{figure}

{An additional dynamical scenario that we considered
involves the evolution of the K-AK pairs in a similar fashion to what was done in Fig.~\ref{set_6}. I.e.,
two localized configurations were
considered, one narrower and another one wider than the stationary (and unstable) K-AK pair, initializing
the system with a symmetric, stationary such pair. As can be seen in Fig.~\ref{set_7}, where the same initial condition was taken, for $\beta<2$, the dynamics is similar to the $\beta=2$ case, including a dissipative evolution due to the role of
the Caputo time-derivative. That is, the generated breather for narrow initial conditions has a vanishing amplitude whereas the wide initial condition leads to an increasing separation (but decelerated in time) for the K-AK pair.}

{Finally, we include in this section the dynamics of breathers.
This aspect involves revisiting for longer time
evolutions the results of~\cite{maciasbountis2022}
in order to determine the asymptotic fate of such
configurations.
To that purpose, we consider initial conditions even in time, namely} $u(x,0)=\phi_\mathrm{br}(x,0)$:
\begin{equation}
    \phi(x,0)=4\arctan\left(\frac{\sqrt{1-\omega ^2}}{\omega\cosh(\sqrt{1-\omega^2}x)}\right),
\end{equation}
and $\dot{u}(x,0)=0$.
For values of $\beta<2$, the oscillations of the breather decay with time. For $u(0,t)$, we fit the maxima and minima to an exponential of the type $ae^{-bt}$ and we consider the characteristic decay time denoted by $\tau=1/b$.

\begin{figure}[h]
\centering
\includegraphics[clip,width=0.45\textwidth,trim=0cm 0cm 0cm 0cm]{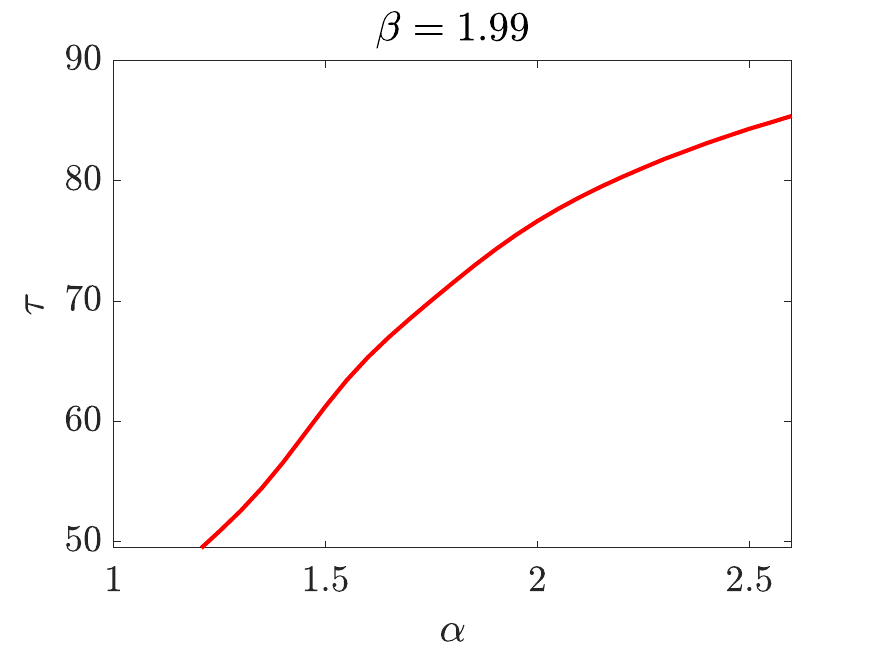}
\includegraphics[clip,width=0.45\textwidth,trim=0cm 0cm 0cm 0cm]{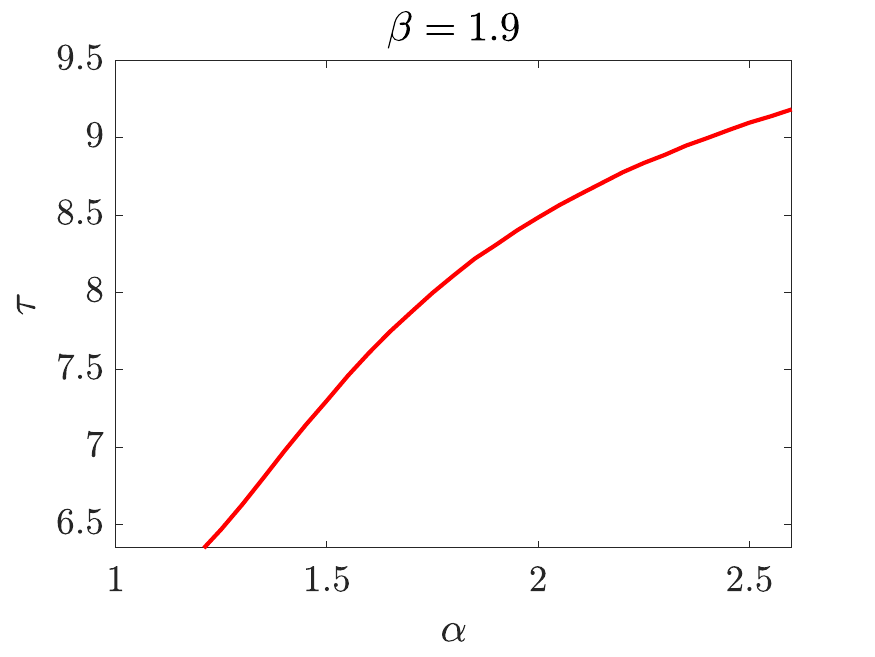}
\caption{Characteristic decay time $\tau$ for $\beta=1.99$ (left panel) and $\beta=1.9$ (right panel) and different values of the Riesz derivative order $\alpha$. Here, we consider $\omega=0.5$.
\label{fig:riesz_caputo_1}}
\end{figure}

In Fig.~\ref{fig:riesz_caputo_1}, we plot $\tau$ for different values of the Riesz derivative order, fixing the Caputo derivative order to $\beta=1.99$ and $\beta=1.9$. The Caputo derivative causes the breather oscillations to decay, as we have previously commented. In this scenario, the Riesz derivative modifies the duration of the transient oscillations, with the oscillations decaying more slowly for higher values of $\alpha$ (for a given $\beta$). Notice the difference in the scale of the vertical axis caused by the difference in the values of the Caputo derivative order. The effect of the Riesz derivative is more dramatic for values of $\beta$ closer to $2$. Also, as is transparent already from earlier sections, for a given $\alpha$, the decay time is higher the closer $\beta$ is to $2$. For values of $\beta$ much smaller than the ones considered here, oscillations are more rapidly damped and $\tau$ is less straightforward to estimate.

\begin{figure}[h]
\centering
\includegraphics[clip,width=0.5\textwidth,trim=0cm 0cm 0cm 0cm]{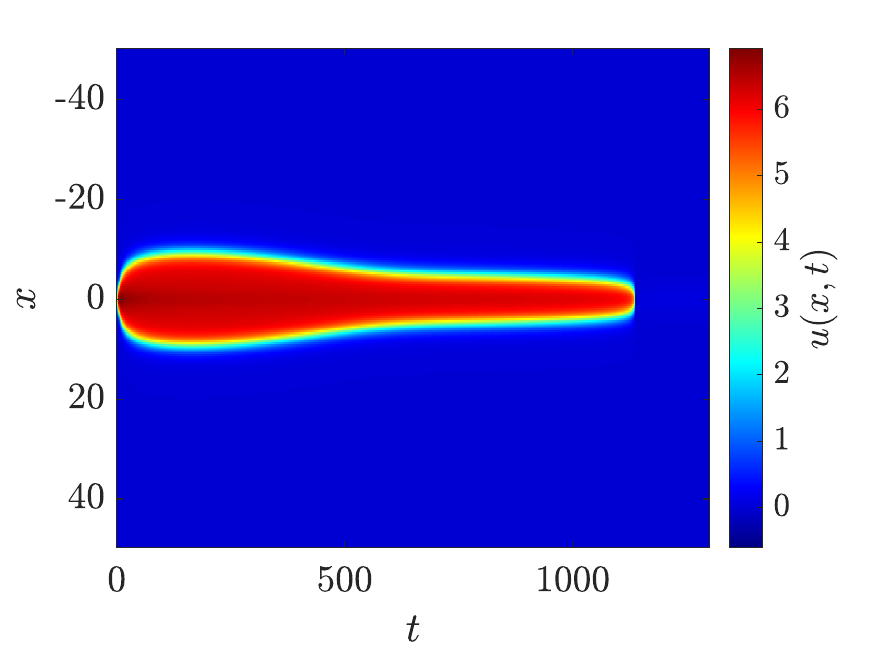}
\caption{Evolution of $u(x,t)$ for $\beta=1.4$, $\alpha=1.8$ and $u_0=0$.
\label{fig:riesz_caputo_3}}
\end{figure}

Finally, it is interesting to consider here a connection with the earlier work of~\cite{maciasbountis2022}, where a different realization of the breather solution was brought to bear.
In particular, in that case, {an odd in time initial condition} $u(x,0)=0$ and
\begin{equation}
    \dot{u}(x,0)=4\sqrt{1-\omega^2}\cdot \sech{(x\sqrt{1-\omega^2)}}
\end{equation}
was considered.
Remarkably, what was found there was that
the system settles in a new transient state, i.e., the end
result of the simulation was fundamentally different than
the previous decaying state and led to the formation of a
K-AK pair.
Indeed, in some cases, the latter state appeared in
the simulations of~\cite{maciasbountis2022} as permanent, however
this was the case due to the time horizon of the simulations
considered.
In Fig.~\ref{fig:riesz_caputo_3}, to better understand this transient state, we show $u(x,t)$ when the initial condition $u(x,0)=0$ is considered. Notice that this transient occurs for a small enough value of the breather frequency $\omega$.

\begin{figure}[h]
\centering
\includegraphics[clip,width=0.45\textwidth,trim=0cm 0cm 0cm 0cm]{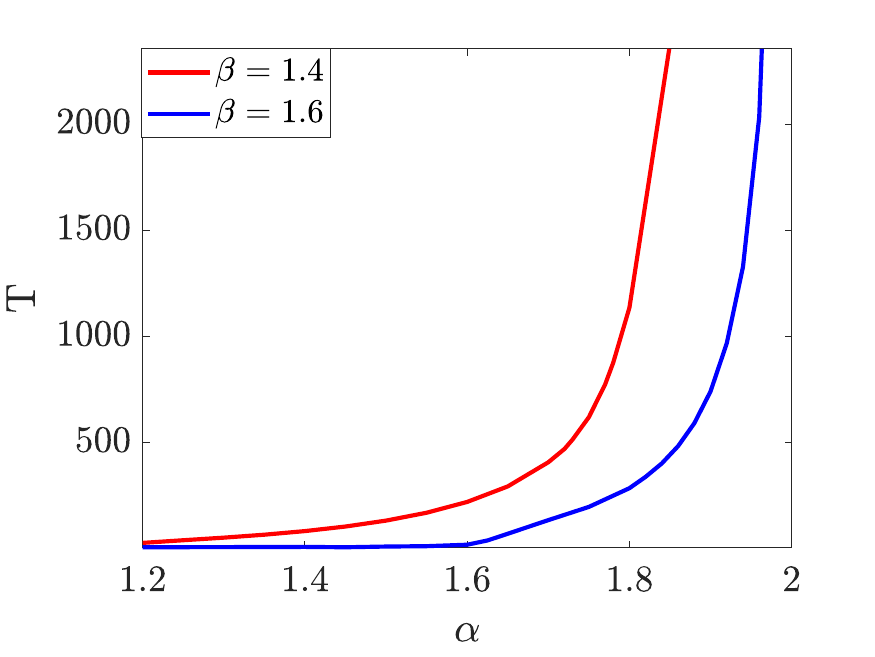}
\includegraphics[clip,width=0.45\textwidth,trim=0cm 0cm 0cm 0cm]{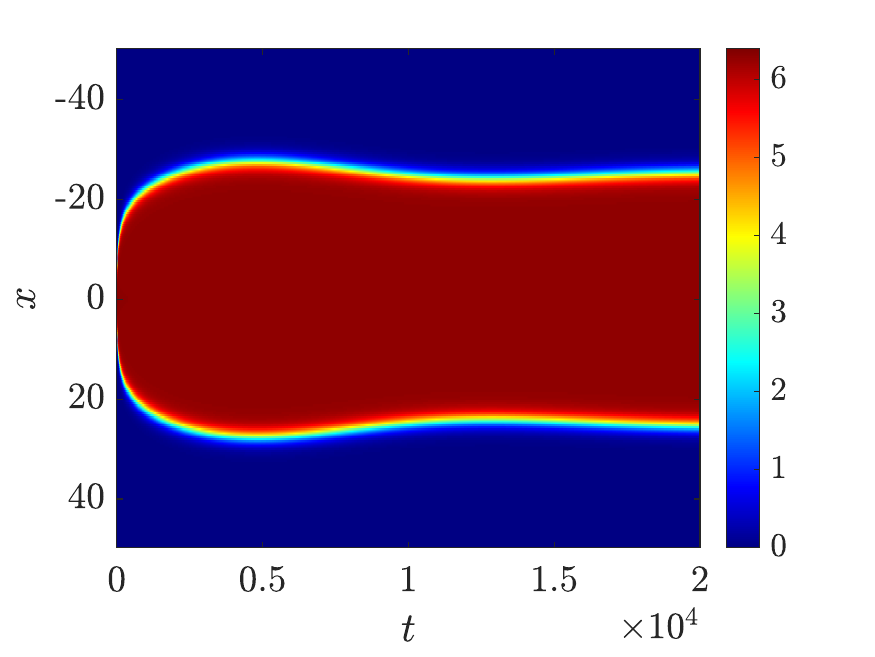}
\caption{Left panel: Duration of the transient bound state dependence on the value of the Riesz derivative for $\beta=1.4$ (red line) and $\beta=1.6$ (blue line). Right panel: Bound state for $\beta=1.7$ and $\alpha=2.01$. As $\alpha\geq 2 $, this state is no longer transient.
\label{fig:riesz_caputo_4}}
\end{figure}

Let us study now in more detail this transient bound state. To this end, we consider variations of $\beta$ and study its time duration, which we name $T$, to avoid confusion with $\tau$ {considered previously for the damping of the kink velocity}. For values of $\beta$ very close to $2$, the behavior of the system is very similar to the non-fractional case.
However, for smaller values of $\beta$, like $\beta=1.4$, this transient  state is observed for a wide range of values of $\alpha$.
In Fig.~\ref{fig:riesz_caputo_4} we show how the duration of the transient bound state depends on the value of the Riesz derivative for $\beta=1.4$. We observe that as $\alpha$ approaches $2$, $T$ increases exponentially. Hence, the relevant state may,
depending on the combination of $\alpha$ and $\beta$, become
extremely long-lived, yet it always has a finite lifetime.

For values of $\alpha \geq 2$ and $1<\beta<2$ this bound state is no longer transient, meaning that the system remains in this state. In the right panel of Fig.\ref{fig:riesz_caputo_4} a simulation with a large final time ($t_{f}=20\cdot 10^{3}$) is shown for $\alpha=2.01$. {In fact, the kink and anti-kink emerging from the breather separate to sufficiently large distances (larger than the equilibrium separation for this $\alpha$), repelling each other. Because of the periodic boundary conditions that naturally stem from the Riesz derivative and the long-range (power-law) nature of the spatial fractional derivative, each member of the pair experiences two repulsive forces, one from its front and another one from its back, so that they equilibrate at $x=\pm L/4$, $L=100$ being the domain length in our case, and consequently the kink/anti-kink eventually stop at that point. We have checked this fact varying many parameters, such as the initial frequency of the breather, values of $\alpha>2$ or $\beta<2$ or domain lengths, resulting in the same outcome. This settling of the kinks at $x=\pm L/4$ was also observed in the above mentioned case of kink-anti-kink interactions and the dynamics of unstable K-AK bound states.}

\section{Conclusions \& Future Challenges}

In the present work, we have explored the fractional
dispersive dynamics of the sine-Gordon equation. We have
done so by considering three independent scenarios.
In the first one, we imposed a Caputo temporal
derivative, while preserving the spatial part of the model.
In the second one, we fixed the temporal second derivative,
and explored the role of variation of the spatial Riesz
derivative away from the Laplacian limit. Finally, having
understood the role of the previous two variations, we
combine them in the final part of our study. In each one
of these cases, we have explored both the dynamics
of a single coherent structure and those of a pair (or
a bound state) of such coherent structures.

Our findings clearly showcase some common conclusion paths.
The role of the Caputo derivative is to induce dissipative
effects in the case where $\beta < 2$. This leads to the
slowdown and eventual stopping of a kink, and may be responsible
for the trapping and eventual elimination of a kink-anti-kink
pair if the latter does not have sufficient kinetic energy.
The presence of a Riesz derivative seems to
maintain an attractive kink-anti-kink interaction
(through the power-law tails of the still monotonic
structures) for $\alpha<2$. On the other hand, it creates the potential for
long-range repulsion through the inter-atomic force for $\alpha>2$.
In the latter case, a saddle equilibrium emerges and the kink
acquires non-monotonic tails. The kink-anti-kink interaction
then evolves in this more complex effective energy landscape
featuring attraction at short and repulsion at long distances.
Finally, the phenomenology of both the Caputo temporal and
the Riesz spatial derivative together showcase both effects,
leading again to decay to stationarity, albeit faster for smaller
$\beta$ and larger $\alpha$.

Naturally, there exist numerous further directions for
future study. On the one hand, while models of the
Klein-Gordon type are extremely useful for building intuition
(lacking the complexity of phase degrees of freedom), it would be
especially instructive to extend considerations to, arguably,
the most widespread dispersive nonlinear PDE model, namely
the nonlinear Schr{\"o}dinger equation; see, e.g.,~\cite{Sulem,AblowitzPrinariTrubatch}. In that case,
we expect some notable differences, such as the fact that
dark soliton bound state
pairs may actually feature stable waveforms rather than
unstable ones (as here)
in the presence of Riesz spatial derivatives.
On the other hand, there have been numerous intriguing
proposals for manipulating dispersion in nonlinear
optics~\cite{BlancoRedondoNC2016,TamOL2019}. Potentially,
adapting such proposals to the settings considered herein
would be especially meaningful.
The very recent experiment of~\cite{hoang2024observationfractionalevolutionnonlinear}, albeit in the bright soliton setting, may pave
the wave for a significant surge of activity in this
direction.
On a related vein,
seeking a deeper understanding of the temporally differentiated
scenario and its dissipation, through a variational or similar
description, would also be of particular interest. These are
among the many novel directions (including also higher dimensional
settings and the interplay therein) meriting further investigation.

\section{Funding}
J.E.M.-D.: The present work reports on a set of final results of the research project “Conservative methods for fractional hyperbolic systems: analysis and applications”, funded by the National Council for Science and Technology of Mexico (CONACYT) through grant A1-S-45928. This material is based upon work supported by the U.S. National Science Foundation under the awards PHY-2110030, PHY-2408988 and DMS-2204702 (PGK). J.C.-M. acknowledges support from the EU (FEDER program 2014–2020) through MCIN/AEI/10.13039/501100011033 (under the projects PID2020-112620GB-I00 and PID2022-143120OB-I00). J.C. acknowledges that this work has been supported by the Spanish State Research Agency (AEI) and the European Regional Development Fund (ERDF, EU) under project PID2023-148160NB-I00 (MCIN/AEI/10.13039/ 501100011033).

\vspace{1cm}

\FloatBarrier
\appendix{{\bf Appendix: Stability of static kinks}}

We consider the case when the Caputo derivative in time is combined with the Riesz derivative in space. We will show that the stability of static kinks is independent of the order of the Caputo derivative.

To this aim, let us introduce a perturbation $w(x,t)$ to the static kink given by $u_0(x)$. That is, $u(x,t)=u_0+\varepsilon w(x,t)$ is introduced in the fractional sine-Gordon equation. Consequently, the perturbation $w(x,t)$ fulfills:

\begin{eqnarray}
\frac{\partial^\beta w}{\partial t^\beta}=
\left[\frac{\partial^\alpha}{\partial x^\alpha} - \cos(u_0(x))\right]w
\end{eqnarray}

This equation can be solved by separation of variables

\begin{equation}
w(x,t)=E_\beta(\lambda^2 t)v(x)
\end{equation}
with $E_\beta(z)$ being the Mittag-Leffler function defined as
\begin{equation}
E_\beta(z)=\sum_{k=0}^\infty \frac{z^k}{\Gamma(\beta k+1)},\ \ z\in\mathbb{C},
\end{equation}
{ with $\Gamma$ being Euler's gamma function
$\Gamma(x)=\int_0^{\infty} t^{x-1} e^{-t} dt$}. The Mittag-Leffler function has the property that \cite{Mainardi}

\begin{equation}
\frac{\partial^\beta E_\beta(\lambda^2 t)}{\partial t^\beta}=\lambda^2 E_\beta(\lambda^2 t)
\end{equation}

Thus, the equation for $v(x)$ reads as

\begin{equation}
\lambda^2 v=\left[\frac{\partial^\alpha}{\partial x^\alpha} - \cos(u_0(x))\right]v
\end{equation}

which is the same equation that one must solve for the stability analysis of the static kink in the case of 2nd derivative in time, except for the time dependence of $w(x,t)$, which is different. However, $E_\beta(\lambda^2 t)$ has a monotonically decreasing dependence (if $0<\beta<1$) or a damped oscillatory dependence (if $1<\beta<2$) when $\lambda^2<0$, and a monotonically (exponential) increase when $\lambda^2>0$; the latter property is easily determined by inspecting the definition of the Mittag-Leffler function. Thus, the spectral picture is analogous to the case of $\beta=2$, with the latter bearing a time dependence $\exp(\lambda t)$. In fact, $E_1(z)=\exp(z)$ and $E_2(z)=\cosh(z)$.

\end{document}